\def\be{\begin{equation}}
\def\ee{\end{equation}}
\def\bea{\begin{eqnarray}}
\def\eea{\end{eqnarray}}
\def\a{\alpha}

\def\l{\lambda}

\documentclass[12pt]{article}

\begin{document}

 \title{Directed percolation in two dimensions: \\ An exact solution}
 \vskip 1cm
\author{L. C. Chen, Department of Mathematics\\
 Fu-Jen Catholic University, Hsinchuang , Taipei 24205, Taiwan\\
and \\ F. Y. Wu, Department of Physics \\
Northeastern University, Boston, Massachusetts 02115, U.S.A.}
\date{}
\maketitle

\begin{abstract}
We consider a directed percolation process on an ${\cal
M}\times {\cal N}$
 rectangular lattice whose vertical edges are directed upward with an occupation
probability $y$  and horizontal edges  directed toward the right
with occupation probabilities $x$ and $1$ in alternate rows. We
deduce a closed-form expression for the percolation probability
$P(x,y)$, the probability that one or more directed paths connect
the lower-left and upper-right corner sites of the lattice.  It is
shown that $P(x,y)$ is critical in the aspect ratio $\a = {\cal
M}/{\cal N}$ at a value $\a_c(x,y) =[1-y^2-x(1-y)^2]/2y^2$ where
$P(x,y)$ is discontinuous, and the critical exponent of the
correlation length for $\a < \a_c(x,y)$ is $\nu=2$.

\end{abstract}




\bigskip

 An outstanding unsolved problem in  stochastic  processes
 is the consideration of directed percolation \cite{durrent,hughes}.
Directed percolation is a Markovian bond percolation process in
which bonds are directed such that only clusters with a ``flow"
are relevant. Very few exact results of directed percolation  are
known. In 1981 Domany and Kinzel \cite{dk} solved one version of a
directed percolation where the occupation
probability is fixed at unity in one spatial direction of a rectangular lattice.
The problem
was subsequently reformulated and solved as a random walk
by one of us and Stanley \cite{ws}. However, the Domany-Kinzel model
is essentially of a one-dimensional
nature due to the restricted freedom in one spatial direction.
To uncover the genuine nature of a two-dimensional directed percolation
it is necessary to relax this uni-directional
restriction.

\medskip
As a first step toward this goal we consider in this paper a
directed percolation
 in which the unity percolation probability
occurs in every {\it other} row of   a rectangular lattice.
 We deduce a closed-form
expression for the percolation probability and analyze its
critical properties for large lattices.

\medskip
We first describe our model. Consider a  2-dimensional rectangular
net of $(M+1)\times (2N+1)$ sites with an aspect ratio
\be
\a = M/2N. \label{aspectratio}
\ee
 Number the sites by $(m,n)$
with $m=0,1,\cdots M,\ n=0,1,\cdots 2N$ as shown in Fig. 1.
Consider a bond percolation process on the lattice with
vertical edges occupied with a probability
$p_y=y$ and horizontal edges in the $n$-th row occupied
with a probability \bea
p_x &=& 1,\quad n={\rm odd} \nonumber\\
    &=& x, \quad n={\rm even}.
\eea

\medskip
Direct edges in the upward direction and toward the right.
Occupied edges form directed paths if traced along
the arrows. In ensuing discussions we shall refer to percolation configurations as  {\it
bond configurations}. A bond configuration is {\it percolating} if
it contains one or more directed paths connecting the two opposite corner sites
$(0,0)$ and $(M,2N)$. A typical percolating configuration is shown
in Fig. 1.

\medskip
In a  bond configuration there are $n_x$ ({\it resp.}  $MN-n_x$)
occupied ({\it resp.}  empty) horizontal edges, and $n_y$
({\it resp.}  $2(M+1)N-n_y$) occupied ({\it resp.}  empty)
vertical edges. Then the {\it percolation probability},
the probability that a bond configuration is
percolating, is
  \be
P_{M,2N}(x,y) =\sum_{perc\  conf} x^{n_x}(1-x)^{MN-n_x} y^{n_y}(1-y)^{2(M+1)N-n_y} \label{PP}
\ee
where the summation is restricted to percolating bond configurations. It is clear that
$0 \leq P_{M,2N}(x,y) \leq 1$ since the summation (\ref{PP}) is identically  $1$
if unrestricted.  It is also clear that $ P_{\infty,2N}(x,y) = 1$ and $ P_{0,2N}(x,y) =0$.
Our interest is to investigate how does $P$ change from $1$ to $0$ as $\a$ varies, and
whether the change is a sharp transition.

\medskip
We state the main result as a Proposition:

\medskip
\noindent
{\it Proposition:

\medskip
For any $x\in[0,1]$ and  $y\in(0,1)$, there exists a critical
aspect ratio
\be
\alpha_c(x,y) =[1-y^2 - x(1-y)^2]/2\, y^2 \label{criticalalpha}
\ee
such that
\begin{eqnarray}
\lim_{N\rightarrow\infty} P_{2\alpha N,2N}(x,y)=\left\{ \begin{array}{lll}
1 & \mbox{ if }\ \alpha>\alpha_c(x,y) \\
0 & \mbox{ if }\ \alpha<\alpha_c(x,y)\\
\frac 12 & \mbox{ if }\ \alpha=\alpha_c(x,y).
 \end{array}
 \right.\label{Proposition}
\end{eqnarray}
Moreover, for $\alpha <\alpha_c(x,y)$, we have the asymptotic behavior
\be
P(2\alpha N,2N)\sim e^{-2N/\xi} \label{correlation}
\ee
where
\be
\xi \sim (\alpha_c-\alpha)^{-2} \label{xi}.
\ee
 }

\medskip
\noindent
 Remarks:

\medskip
1. Equation (\ref{correlation}) defines $\xi$ as
the correlation length  and Eq. (\ref{xi}) gives the
correlation length
critical exponent $\nu = 2$.

\medskip
2. For $x=1$ our model reduces to the Domany-Kinzel model
\cite{dk,ws} on an $(M+1)\times (2N+1)$ lattice and
(\ref{criticalalpha}) leads to $\a_c = (1-y)/y$ in agreement with
previous result. For $x=0$ our model is  again a Domany-Kinzel model
but on an $(M+1) \times (N+1)$ lattice with a vertical edge
occupation probability $y^2$. Our result gives the critical aspect ratio
$2\, \a_c = (1-y^2)/
y^2$ again  in agreement with \cite{dk,ws}.

  \medskip\noindent
{\it Proof of the Proposition}:

\medskip
The main body of this paper is the proof of the Proposition.

\medskip
There are $2N$ rows of vertical edges in the lattice.
 Number these rows from $1$ to $2N$ starting from the bottom. An occupied
vertical edge in a bond configuration is {\it wet} if it  lies on
a percolating path connecting $(0,0)$ and $(M,2N)$, and is  {\it
primary} wet if it is the first wet edge (in a row of vertical
edges) counting from the left.
In the bottom row of vertical edges in
Fig. 1, for example, there are two wet edges and
the primary wet edge is the one
connecting sites $(1,0)$ and $(1,1)$. In a percolating configuration
there is one primary wet edge
 in every row and these edges carry an overall occupation probability   $y^{2N}$.
Since a bond configuration is percolating whenever a vertical edge in the
$2N$-th row  is primary wet, which can occur at any of
the $m$-th horizontal positions $m=0,1,\cdots,M$, we have
  \be
P_{M,2N}(x,y) = y^{2N} \sum _{m=0}^M w_{m, 2N}.\label{P}
 \ee
Here $y^{2n} w_{m,2n}$ is the probability that
the primary
 wet edge in the $(2n)$-th
row occurs  at the horizontal position $m$.

\medskip
We first establish a Lemma:

\medskip
\noindent
{\it Lemma:
\begin{equation}
w_{m,2n} = \frac 1 {2\pi i} \oint \frac {dt}{t^{m+1}(1-at+bt^2)^n}
\label{lemma}
\end{equation}
where the contour of integration is around the unit circle and
\begin{equation}
a= 1 - y^2 + x(1-y)^2, \quad \quad b = x(1-y)^2.\label{ab}
\end{equation}

\noindent
Proof of the Lemma:}

\medskip
 It is not difficult to see that the function $w_{m,2n}(x,y)$ satisfies the recursion relation
\be
w_{m,2n}=\sum _{k=0}^m w_{k,2}w_{m-k,2n-2} \label{recursion}
\ee
and the initial condition
 \be
w_{m,0} = \delta_{\rm Kr}(m,0). \label{bc}
\ee

\medskip
Define  generating functions
\bea
W_1(t) &=& \sum_{m=0}^\infty w_{m,2} t^m \label{F} \\
W_2(t,s) &=& \sum_{m=0}^\infty \sum_{n=0}^\infty
         w_{m, 2n} t^m s^n .
  \label{G}
\eea
 Substituting (\ref{recursion}) into (\ref{G}) and
 changing the order of summation by using
$\sum_{m=0}^\infty \sum_{n=0}^m = \sum_{n=0}^\infty\sum_{m=n}^\infty $,
 we obtain after some rearrangement and the use of (\ref{bc}),
\bea
W_2(t,s)
 = 1+ s\, W_1(t) W_2(t,s) \nonumber
\eea
which yields
\be
W_2(t,s) = \frac 1 {1-sW_1(t)}. \label{gen}
\ee
We can now invert (\ref{G}) to obtain
\bea
w_{m,2n}  &=&\frac 1 {(2\pi i)^2} \oint \frac {dt} { t^{n+1} }
\oint  \frac {ds} { s^{n+1} } \bigg(  \frac 1 {1-sW_1(t)}\bigg) \nonumber \\
    &=& \frac 1 {2\pi i} \oint \frac {dt }{t^{m+1}}  \Big[W_1(t)\Big]^n, \label{m2n}
\eea
where the contour of integration is around the unit circle.

\medskip
To compute $W_1(t)$
we need to evaluate $w_{m,2}(x,y)$ for an $(m+1)\times 3$ lattice.
 There are now 2 rows of vertical edges.
As aforementioned $y^2w_{m,2}$ is the probability
that $(0, 0)$ is connected to $(m, 2)$ with  the primary
wet vertical edge in the top row  occurring at $m$.
However the primary wet vertical edge in the bottom row can be at any
$j$ in $0\leq j \leq m$.  Denote the probability for this to occur by
$y^2  \l_j (1-y)^{m-j} x^{m-j} $.
  Then we have
 \be
w_{m,2} = \sum _{j=0}^m \l_j (1-y)^{m-j} x^{m-j}, \label{gh}
\ee
where the factor $(1-y)^{m-j} x^{m-j}$
ensures that the primary wet edge in the top row is at $m$
as shown  in Fig 2(a).
  Particularly, we have $w_{0,2} = \lambda_0 =1$.
 
\medskip
The factor $\l_j$ in (\ref{gh}) satisfies a
 recursive relation which can be written as
\be
\l_{j} = (1-y) \l_{j-1} + y(1-x)(1-y)\, w_{j-1,2},
\quad j=1,2,\cdots, m. \label{lambda}
\ee
The two terms on the right-hand side of (\ref{lambda}) arise from
the two possibilities that the vertical
edge connecting $(j-1,0)$ and $(j-1,1)$ is either empty  (with probability $1-y$)
 or occupied (with probability $y$) as shown in the two panels in Fig. 2(b).
 In the latter case
the factor $(1-x)(1-y)$
 ensures that the site $(j-1,1)$ is not on a percolating path.

\medskip
To solve the coupled recursion relations (\ref{gh})
and (\ref{lambda}),
define the generating function
\bea
\Lambda (t) = \sum _{j=0} ^\infty \lambda_j \,t^j. \nonumber
\eea
Multiplying (\ref{gh}) and (\ref{lambda}) by $t^m$ and $t^{j-1}$, respectively,
and summing over
 $m$ and $j-1$ from $0$ to $\infty$, we obtain after some manipulation
 \bea
W_1(t)  &=&\frac {y^2 \Lambda(t)} {1-x(1-y)t},\nonumber \\
\frac 1 t \Bigg( {\Lambda(t) -1} \Bigg) &=& (1-y) \Lambda(t) +y(1-x)(1-y) W_1(t) .
\eea
This gives
 \be
W_1(t)  = \frac 1 {1-at+bt^2} \label{W1}
\ee
after eliminating $ \Lambda (t)$
where $a,b$ are given in (\ref{ab}).
The substitution of (\ref{W1})
 into (\ref{m2n}) establishes the Lemma.

\medskip
We now continue the proof of the Proposition.

\medskip
Substitute (\ref{lemma}) into (\ref{P}) and carry out the
summation in $m$.  This leads to
 \begin{equation}
  P_{M,2N}(x,y)= \frac{y^{2N}} {2\pi i} \oint_{C+}
  \frac {dt} {(t-1)(1-a\,t+b\,t^2)^N} \bigg(1- \frac 1 {t^{M+1}}\bigg)
  \label{P1}
 \end{equation}
where the contour $C+$ is a  circle
enclosing the unit circle. Let $t_1$ and $t_2$ be
 the two roots of
$1-at+bt^2=0$, both of which are real. We have $t_1t_2 = 1 /b, \, t_1+t_2= a/b$,
and hence
\bea
(t_1-1) (t_2-1) &=& t_1t_2 -(t_1+t_2) +1 \nonumber \\
                &=& \frac 1 b - \frac a b +1 = \frac y {(1-x)^2} > 0, \nonumber
\eea
so both  $t_1$ and $t_2$ lie outside the unit circle.  We can therefore
choose the radius of $C+$ to be greater than $1$ but
smaller than both $t_1$ and $t_2$ so that $C+$ encloses only the simple pole  $t=1$
in (\ref{P1}).
It follows that the
first term on the right-hand side of (\ref{P1}) picks up only the
residue at $t=1$ which is
\bea
\frac {y^{2N}} {(1-a+b)^N} =1, \nonumber
\eea
  and we obtain
 \bea
 P_{M,2N}(x,y)=1-I_{M,N} \nonumber
 \eea
 where
 \begin{equation}
 I_{M,N} =  \frac{y^{2N}} {2\pi i} \oint_{C+}
 \frac {dt} {(t-1)t^{M+1}(1-at+bt^2)^N}.
 \label{I}
 \end{equation}
 Note that since $|t|>1$ along $C+$ (\ref{I}) leads to the expected result
 $ P_{\infty,2N}=1$.

\medskip
To further evaluate $I_{M,N}$ we introduce $z=1/t$ to write
\be
I_{M,N}=\frac{y^{2N}} {2\pi i} \oint_{C-}
 \frac {z^{M+2N}dz} {(z-1)(z^2-az+b)^N}
 \label{I1}
\ee
where the contour $C-$ is now within the unit circle.

\medskip
For $M,N$ large and fixed aspect ratio $\alpha = M/2N$,
we can rewrite (\ref{I1}) as
\be
I_{M,N}=\frac 1 {2\pi i} \oint_{C-}
 \frac {dz} {z-1} \Big[ f_\a (z) \Big]^N
 \label{I2}
\ee
where
\bea
f_\a(z) = \frac {y^2 z^{2+\a}} {z^2 -az +b}. \nonumber
\eea
The integral $I_{M,N}$  can be evaluated using the method of steepest descent
\cite{morse,dkbook}
by deforming the contour to pass a point $z=z_0$ where $f_\a(z)$  is stationary.
To the leading order this gives $I_{M,N} \sim [f_\a(z_0)]^N$.  Moreover,
since $ I_{M,N} \leq 1$,
we must have
 $ f_\a(z_0) \leq 1$  with the equal sign holding at $f_\a(z_0)=1$.
Thus a transition occurs  at $z_0=1$.

\medskip
Now
\bea
f'_\a(z) = \frac {y^2 z^{1+\a}} {(z^2-az+b)^2} \bigg[ \a z^2 - (1+\a)
 a\,z + (2+\a)b \bigg] \nonumber
\eea
and the stationary point $z_0$ is determined by
 \bea
 \a z_0^2 - (1+\a) a\,z_0 + (2+\a)b=0. \nonumber
\eea

\medskip
The critical condition $z_0=1$ now gives
 \be
 \a = \frac {a-2b}{1-a+b} = \a_c (x,y) \label{criticalfrontier}
\ee
where $\a_c(x,y)$ is given in (\ref{criticalalpha}).
It is readily verified that we have $(d \a / d z) _{z=1} <0$ along
(\ref{criticalfrontier}).
 Thus, for $\a>\a_c(x,y)$, the stationary point $z_0$ lies within the unit circle
so we can deform $C-$ continuously to pass $z_0$,
and obtain $I_{2\a N,N} = [ f_\a(z_0)]^N \sim 0$.
This gives $P_{2\a N, N} (x,y) \sim 1$ which establishes the first line
of (\ref{Proposition}).

\medskip
On the other hand, for $\a<\a_c(x,y)$, $z_0$ occurs outside the unit circle
and when the contour $C-$ is deformed  to pass $z_0$ it must cross the simple
pole at $z=1$ and picks up the residue at the pole, which is equal to $1$.
This gives $I_{2\a N,N} \sim 1 - [ f_\a(z_0)]^N $
and   $P_{2\a N,N} (x,y) \sim [ f_\a(z_0)]^N \sim  0 $ for large $N$.
 This establishes the second line of (\ref{Proposition}).

\medskip
For $\a=\a_c(x,y)$, $z_0$ is on the unit circle so the crossing of the contour
at $z=1$ picks up only half of the residue, namely, $1/2$.
This establishes the third line of (\ref{Proposition}).

\medskip
Finally, for $\a<\a_c(x,y)$,
the method of steepest decent \cite{morse,dkbook} dictates that we have
\bea
\big[f_\a(z_0)\big]^N = e^{N\ln [f_\a(z_0)]} \sim
 e^{-NC_1(x,y) (z_0-1)^2} \sim e^{- N C_2(x,y)(\a-\a_c)^2} \nonumber
\eea
where expressions of $C_1(x,y)$ and $C_2(x,y)$,
which do not affect our conclusions, can be explicitly evaluated.
 This establishes the asymptotic behavior (\ref{correlation}) with $\xi = 2/[C_2(x,y)(\a-\a_c)^2]$.

\medskip
We have completed the proof of the Proposition.

\medskip
In summary, we have obtained a closed-form expression for the percolation
probability $P_{M,2N}(x,y)$ for the directed percolation process in which
the occupation probability is $y$ in the vertical direction and alternately
$x$ and $1$ in the horizontal direction.  For $M,N$ large, the percolation
probability exhibits a critical behavior at $\a = \a_c$.  The correlation
length $\xi$ for $\a < \a_c$ is found to diverge with the critical exponent $\nu=2$.
While these properties are similar to those found in the Domany-Kinzel
model \cite{dk,ws}, our analysis permits the relaxation of
 the restriction of unit occupation
probability in one spatial direction.
It is hoped that the analysis serves as the first step of further relaxation
in percolation probabilities, eventually leading
to an understanding of genuine 2-dimensional directed percolation processes.

\medskip
This work was initiated while FYW was at the National Center of
Theoretical Sciences (NCTS) in Taipei.  The support of the NCTS is
gratefully acknowledged. Work of LCC has been supported in part
by the National Science Council, Taiwan.

\bigskip

\centerline{\bf Figure Captions}

\vskip 1cm

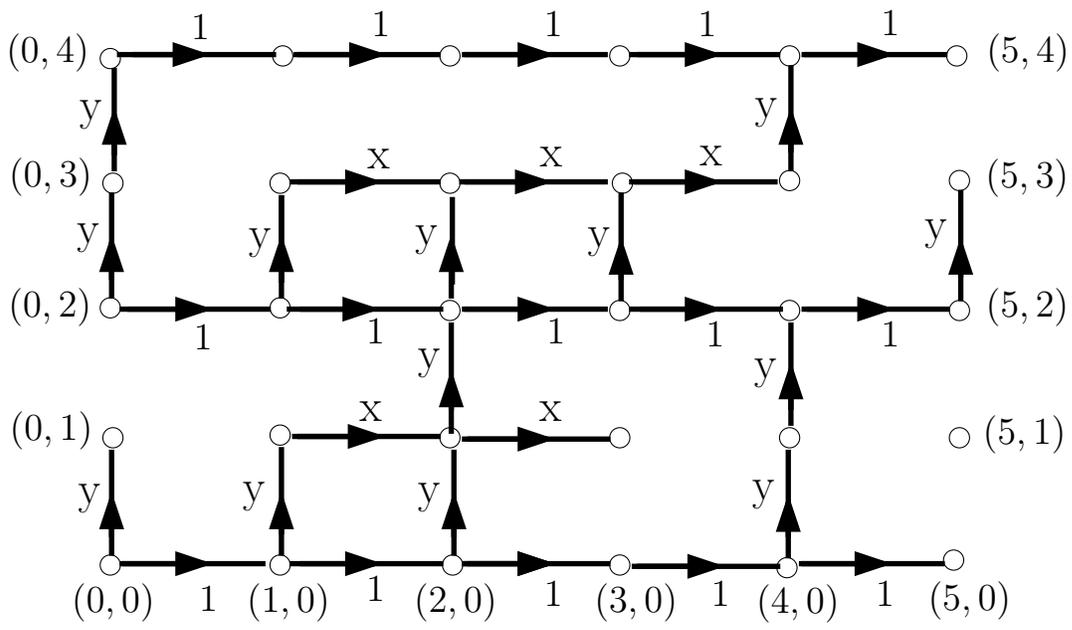
Fig. 1. A typical percolating  configuration on a $6\times 5$ lattice ($M=5, N=2$).
Open circles denote lattice sites.
Oriented edges are occupied with weights shown. Empty edges carry weights $1-x$
and $1-y$ in horizontal and vertical directions respectively.

\vskip .5cm
Fig. 2. Construction of recursion relations.  (a) Construction of (\ref{gh}).
(b) Construction of (\ref{lambda}).
Occupied edges are shown as oriented edges;
dotted edges can be either occupied or empty. To
each row of vertical edges there is
an additional factor $y$ not shown.

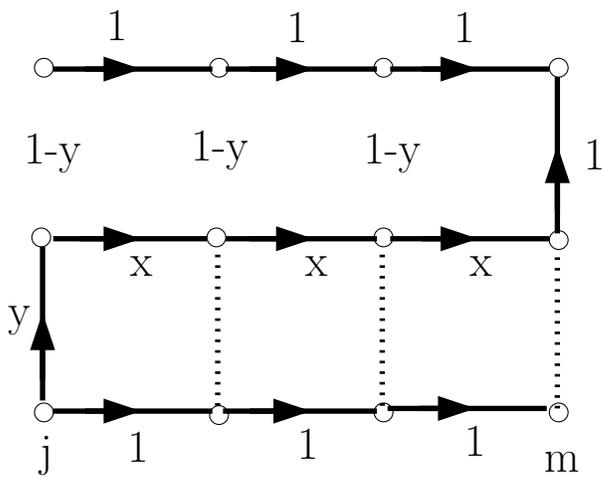
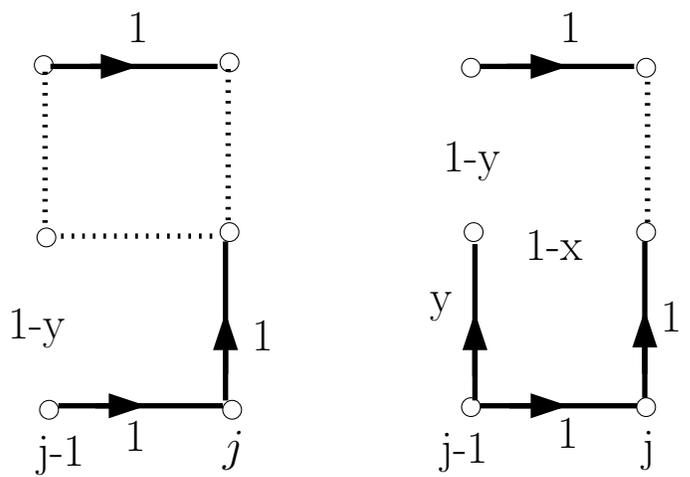
\end{document}